\documentclass{article}
\usepackage{spconf,amsmath,graphicx}
\usepackage{hyperref}
\usepackage{url}
\usepackage[utf8]{inputenc} 
\usepackage[T1]{fontenc}    
\usepackage{booktabs}       
\usepackage{amsfonts}       
\usepackage{nicefrac}       
\usepackage{microtype}      
\usepackage{xcolor}         
\usepackage{amsmath}
\usepackage{amssymb}
\usepackage{mathtools}
\usepackage{makecell}
\usepackage{caption}
\usepackage{stackrel}
\usepackage{enumitem}
\usepackage{multirow}

\definecolor{Red}{rgb}{0.768, 0.054, 0.054}
\definecolor{Blue}{rgb}{0.152, 0.294, 0.925}
\definecolor{Green}{rgb}{0,0.4,0.7}
\hypersetup{
    colorlinks=true,
    citecolor=teal,
    linkcolor=Red,
    urlcolor=Green,
}
\definecolor{red}{RGB}{205,33,42}

\usepackage{amsmath,amsfonts,bm}

\def\eqref#1{equation~\ref{#1}}

\def\1{\bm{1}}

\def\vs{{\bm{s}}}

\def\vx{{\bm{x}}}
\def\vy{{\bm{y}}}

\def\mH{{\bm{H}}}
\def\mI{{\bm{I}}}

\def\mW{{\bm{W}}}
\def\mX{{\bm{X}}}
\def\mY{{\bm{Y}}}

\DeclareMathAlphabet{\mathsfit}{\encodingdefault}{\sfdefault}{m}{sl}
\SetMathAlphabet{\mathsfit}{bold}{\encodingdefault}{\sfdefault}{bx}{n}

\def\sR{{\mathbb{R}}}

\title{Grad-StyleSpeech: Any-speaker adaptive \\ Text-to-Speech Synthesis with Diffusion Models}
\name{Minki Kang$^{1,2*}\thanks{*: Equal Contribution}$, Dongchan Min$^{1,2*}$, Sung Ju Hwang$^{1,2}$}
\address{AITRICS$^1$, KAIST$^2$ \\
\small\texttt{{zzxc1133@aitrics.com, \quad \{alsehdcks95, sjhwang82\}@kaist.ac.kr}}}

\begin{document}
%
\maketitle

\begin{abstract}
There has been a significant progress in Text-To-Speech (TTS) synthesis technology in recent years, thanks to the advancement in neural generative modeling.
However, existing methods on any-speaker adaptive TTS have achieved unsatisfactory performance, due to their suboptimal accuracy in mimicking the target speakers' styles.
In this work, we present Grad-StyleSpeech, which is an any-speaker adaptive TTS framework that is based on a diffusion model that can generate highly natural speech with extremely high similarity to target speakers' voice, given a few seconds of reference speech. Grad-StyleSpeech significantly outperforms recent speaker-adaptive TTS baselines on English benchmarks.
Audio samples are available at \href{https://nardien.github.io/grad-stylespeech-demo}{https://nardien.github.io/grad-stylespeech-demo}.
\end{abstract}
\section{introduction}

Recently, the deep neural network-based Text-To-Speech (TTS) synthesis models have shown remarkable performance on both quality and speed, thanks to the progress on the generative modeling~\cite{Transformer, Glow-TTS}, non-autoregressive acoustic models~\cite{FastSpeech2}, and the powerful neural vocoder~\cite{HifiGAN}.
Remarkably, diffusion models~\cite{DDPM, SDE} have recently been shown to synthesize high-quality images in the image generation tasks and speech in the TTS synthesis tasks~\cite{DiffTTS, GradTTS, Guided-TTS}.
Beyond the TTS synthesis on a single speaker, recent works~\cite{VITS, MultiSpeech} have shown decent quality in synthesizing the speech of multiple speakers.
Furthermore, a variety of works~\cite{NeuralVoiceCloning,SEA,SpeakerAdaptation,StyleSpeech,AdaSpeech,YourTTS} focus on \textbf{any-speaker adaptive} TTS where the system can synthesize the speech of any speaker given the reference speech of them.
Due to its extensive possible applications in the real world, the research on any-speaker adaptive TTS -- sometimes, termed Voice Cloning -- has grown and highlighted a lot.

Most of works on any-speaker adaptive TTS focus to synthesize the speech which is highly natural and similar to the target speaker's voice, given the few samples of speech from the target speaker.
Some of previous works~\cite{SEA, AdaSpeech} need a few transcribed (supervised) samples for fine-tuning TTS model. They have a clear drawback where they require the supervised samples from the target speaker and heavy computational costs to update the model parameters.
On the other hand, recent works~\cite{NeuralVoiceCloning, SpeakerAdaptation, StyleSpeech} have focused on a zero-shot approach in which transcribed samples and the additional fine-tuning stage are not necessarily required to adapt to the unseen speaker, thanks to their neural encoder that encode any speech into the latent vector.
However, due to their capacity for generative modeling~\cite{oversmoothness}, they frequently show lower similarity on unseen speakers and are vulnerable to generating speech in an unique style, such as emotional speech.

In this work, we propose a zero-shot any-speaker adaptive TTS model, \emph{Grad-StyleSpeech}, that generates highly natural and similar speech given only few seconds of reference speech from any target speaker with a score-based diffusion model~\cite{SDE}. 
In contrast to previous diffusion-based approaches~\cite{DiffTTS, GradTTS, Guided-TTS}, we construct the model using a style-based generative model~\cite{StyleSpeech} to take the target speaker's style into account.
Specifically, we propose a hierarchical transformer encoder to generate a representative prior noise distribution where the reverse diffusion process is able to generate more similar speech of target speakers, by reflecting the target speakers' style during embedding the input phonemes.
In experiments, we empirically show that our method outperforms recent baselines.

Our main contributions can be summarized as follows:
\vspace{-0.1in}
\begin{itemize}[itemsep=0.5mm, parsep=1pt, leftmargin=*]
    \item In this paper, we propose a TTS model \textit{Grad-StyleSpeech} which synthesizes the production quality of speech with any speaker's voice, even with a zero-shot approach.
    \item We introduce a hierarchical transformer encoder to build the representative noise prior distribution for any-speaker adaptive settings with score-based diffusion models. 
    \item Our model outperforms recent any-speaker adaptive TTS baselines on both LibriTTS and VCTK datasets.
\end{itemize}
\section{Method}

\begin{figure}
    \centering
    \includegraphics[width=0.8\linewidth]{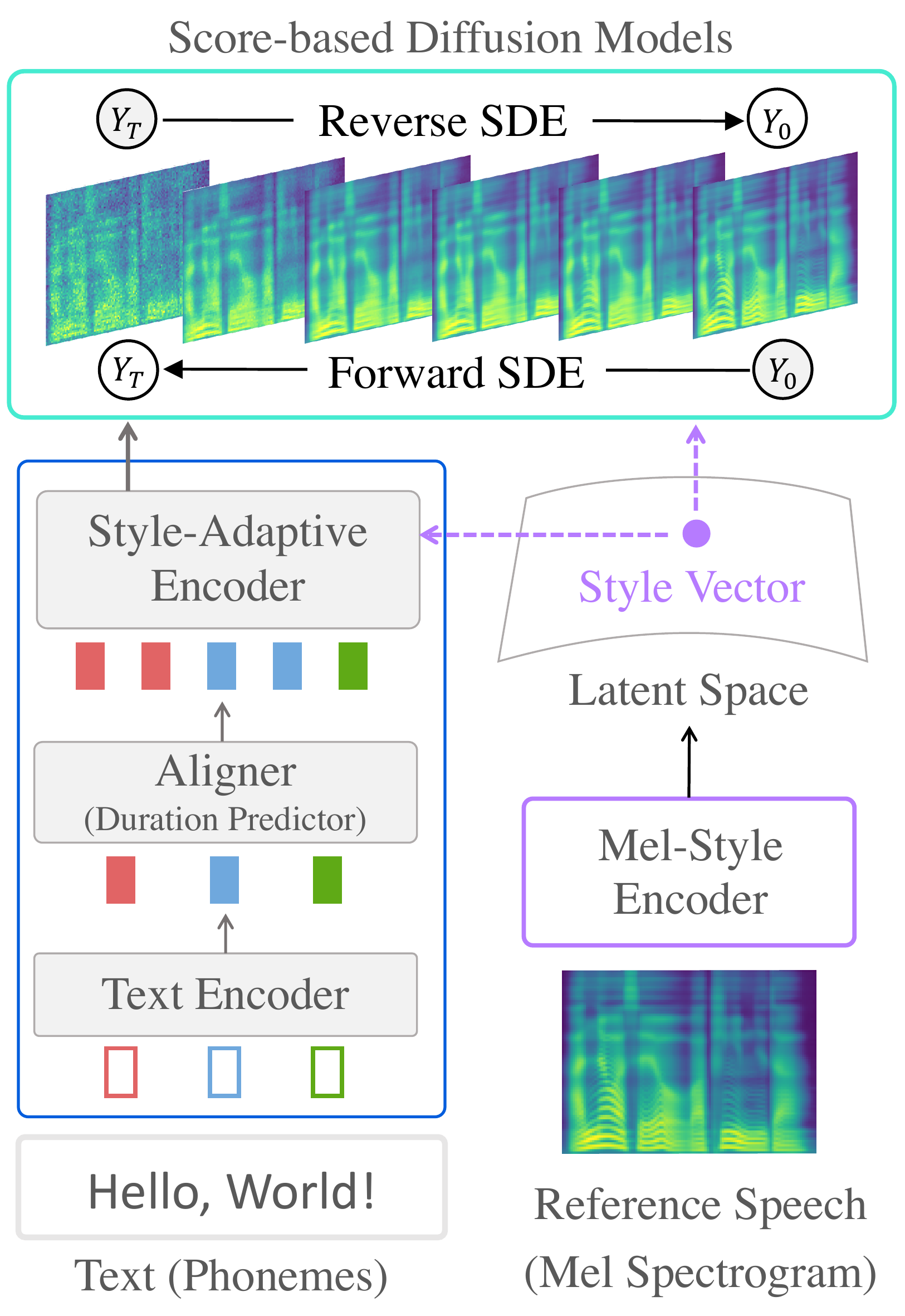}
    \vspace{-0.15in}
    \caption{\textbf{Framework Overview.} Blue box indicates a hierarchical transformer encoder where it outputs $\bm{\mu}$.}
    \vspace{-0.15in}
    \label{fig:model}
\end{figure}

Speaker Adaptive Text-To-Speech (TTS) task aims to generate the speech given the text transcription and the reference speech of the target speaker.
In this work, we focus to synthesize the mel-spectrograms (audio feature) instead of the raw waveform as in previous works~\cite{StyleSpeech,AdaSpeech}.
Formally, given the text $\vx = [x_1, \ldots, x_n]$ consists of the phonemes and the reference speech $\mY = [\vy_1, \ldots, \vy_m] \in \sR^{m \times 80}$, the objective is to generate the ground-truth speech $\tilde{\mY}$. 
In the training stage, $\tilde{\mY}$ is identical with the reference speech $\mY$, while not in the inference stage.

Our model consists of three parts: a mel-style encoder that embeds the reference speech into the style vector~\cite{StyleSpeech}, a hierarchical transformer encoder that generates the representations conditioned on the text and the style vector, and a diffusion model which generates mel-spectrograms by denoising steps~\cite{GradTTS}.
We illustrate our overall framework in Figure~\ref{fig:model}.

\subsection{Mel-Style Encoder}
\label{method:mel-style-encoder}
As a core component for the zero-shot any-speaker adaptive TTS, we use the mel-style encoder~\cite{StyleSpeech} to embed the reference speech into the latent style vector. Formally, $\vs = h_\psi (\mY)$ where $\vs \in \sR^{d'}$ is the style vector and $h_\psi$ is the mel-style encoder parameterized by $\psi$. 
Specifically, the mel-style encoder consists of spectral and temporal processor~\cite{NeuralVoiceCloning}, the transformer layer consisting of the multi-head self-attention~\cite{Transformer}, and the temporal average pooling at the end.

\subsection{Score-based Diffusion Model}
Diffusion model~\cite{DDPM} generates samples by progressively denoising the noise sampled from the prior noise distribution, which is generally a unit gaussian distribution $\mathcal{N}(\boldsymbol{0}, \mI)$.
We mostly follow the formulation introduced in Grad-TTS~\cite{GradTTS}, which defines the denoising process in terms of SDEs instead of Markov chains~\cite{SDE}.
In this subsection, we briefly recap essential parts of the score-based diffusion model.
\subsubsection{Forward Diffusion Process}
The forward diffusion process progressively adds the noise drawn from the noise distribution $\mathcal{N}(\boldsymbol{0}, \mI)$ to the samples drawn from the sample distribution $\mY_0 \sim p_0$. We define the differential equation for forward diffusion process as follows:
\begin{equation*}
    d\mY_t = -\frac{1}{2} \beta(t) \mY_t dt + \sqrt{\beta(t)} d\mW_t,
\end{equation*}
where $t \in [0, T]$ is the continuous time step, $\beta(t)$ is the noise scheduling function, and $\mW_t$ is the standard Wiener process~\cite{SDE}. 
Instead, Grad-TTS~\cite{GradTTS} proposes to gradually denoise the noisy samples from the data-driven prior noise distribution $\mathcal{N}(\bm{\mu}, \mI)$ where $\bm{\mu}$ is the text- and style-conditioned representations from the neural network.
\begin{equation}
\label{eqn:forward-diffusion}
    d\mY_t = - \frac{1}{2} \beta(t) (\mY_t - \bm\mu) dt + \sqrt{\beta(t)} d\mW_t.
\end{equation}
It is tractable to compute the transition kernel $p_{0t}(\mY_t | \mY_0)$ where it is also a Gaussian distribution~\cite{SDE} as follows:
\begin{gather}
\label{eqn:transition}
   p_{0t} (\mY_t | \mY_0) = \mathcal{N} (\mY_t; \bm\gamma_t, \bm\sigma_t^2), \quad \bm\sigma_t^2 = \mI - e^{-\int_0^t \beta(s) ds} \\
   \bm\gamma_t = (\mI - e^{-\frac{1}{2} \int_0^t \beta(s) ds}) \bm\mu + e^{-\frac{1}{2} \int_0^t \beta(s) ds}\mY_0. \nonumber
\end{gather}

\subsubsection{Reverse Diffusion Process}
\label{section:reverse-diffusion}
On the other hand, the reverse diffusion process gradually inverts the noise from $p_T$ into the data samples from $p_0$. Following the results from ~\cite{Anderson} and~\cite{SDE}, reverse diffusion process of Equation~\ref{eqn:forward-diffusion} is given as the reverse-time SDE as follows~\cite{GradTTS}:
\begin{align*}
    d\mY_t &= \left[ -\frac{1}{2} \beta(t) (\mY_t - \bm\mu) - \beta(t) \nabla_{\mY_t} \log p_t(\mY_t) \right] dt \\
           &+ \sqrt{\beta(t)} d\tilde{\mW}_t,
\end{align*}
where $\tilde{\mW}_t$ is a reverse Wiener process and $\nabla_{\mY_t} \log p_t(\mY_t)$ is the score function of the data distribution $p_t(\mY_t)$. We can apply any numerical SDE solvers for solving reverse SDE to generate samples $\mY_0$ from the noise $\mY_T$~\cite{SDE}. Since it is intractable to obtain the exact score during the reverse diffusion process, we estimate the score using the neural network $\bm\epsilon_\theta(\mY_t, t, \boldsymbol{\mu}, \vs)$.

\subsection{Hierarchical Transformer Encoder}
\label{method:text-encoder}
We empirically find that the composition of $\bm\mu$ is important in the multi-speaker TTS with the diffusion model.
Therefore, we compose the encoder into three-level hierarchies. First, the text encoder $f_\lambda$ maps the input text into the hidden representations through the multiple transformer blocks~\cite{Transformer} for the contextual representation of the phoneme sequence: $\mH = f_\lambda(\vx) \in \sR^{n \times d}$. 
Then, we use the unsupervised alignment learning framework~\cite{Align} which computes the alignment between the input text $\vx$ and the target speech $\mY$ and regulate the length of representations after the text encoder to the length of the target speech: $\texttt{Align}(\mH, \vx, \mY) = \tilde{\mH} \in \sR^{m \times d}$.
We also use the duration predictor to predict the duration of each phoneme in the inference time.
Finally, we encode the length-regulated embedding sequence through the style-adaptive transformer blocks to build the speaker-adaptive hidden representations: $\bm{\mu} = g_\phi(\tilde{\mH}, \vs)$ where $\vs$ is the style vector defined in $\S$\ref{method:mel-style-encoder}.
We use the Style-Adaptive Layer Normalization (SALN)~\cite{StyleSpeech}, to condition the style information into the transformer blocks in the style-adaptive encoder.
Consequently, the hierarchical transformer encoder outputs the hidden representations $\bm{\mu}$ that reflect the linguistic contents from the input text $\vx$ and the style information from the style vector $\vs$. 
Above $\bm{\mu}$ is used for the style-conditioned prior noise distribution in the denoising diffusion model described in the previous section.
Following Grad-TTS~\cite{GradTTS}, we add the prior loss $\mathcal{L}_{prior} = \lVert \bm{\mu} - \mY \rVert_2^2,$
where we minimize the L2 distance between the $\bm{\mu}$ and $\mY$.

\subsection{Training}
To train the score estimation network $\bm\epsilon_\theta$ in \S~\ref{section:reverse-diffusion}, we compute the expectation with marginalization over the tractable transition kernel $p_{0t}(\mY_t|\mY_0)$:
\begin{align}
    \mathcal{L}_{diff} = \mathbb{E}_{t \sim \mathcal{U}(0,T)} &\mathbb{E}_{\mY_0 \sim p_0(\mX_0)} \mathbb{E}_{\mY_t \sim p_{0t}(\mY_t|\mY_0) } \\
                &\lVert \bm\epsilon_\theta (\mY_t, t, \boldsymbol{\mu}, \vs) - \nabla_{\mY_t} \log p_{0t}(\mY_t|\mY_0) \rVert^2_2, \nonumber
\end{align}
where $\vs$ is the style vector in \S~\ref{method:mel-style-encoder} and $\mY_t$ is sampled from the Gaussian distribution depicted in Equation~\ref{eqn:transition}. Then, the exact score computation is tractable in this form as follows:
\begin{align}
    \label{eqn:diffusion-loss}
    &\mathcal{L}_{diff} = \\
    &\mathbb{E}_{t \sim \mathcal{U}(0,T)} \mathbb{E}_{\mY_0 \sim p_0(\mY_0)} \mathbb{E}_{\bm\epsilon \sim \mathcal{N}(\boldsymbol{0}, \mI)} \lVert \bm\epsilon_\theta (\mY_t, t, \bm{\mu}, \vs) 
                                            + {\bm\sigma_t}^{-1}{\bm\epsilon} \rVert^2_2, \nonumber
\end{align}
where $\bm\sigma_t = \sqrt{1 - e^{-\int_0^t \beta(s) ds}}$ as defined in Equation~\ref{eqn:forward-diffusion}.

Combining above loss terms including $\mathcal{L}_{align}$ for the aligner and duration predictor trainig~\cite{Align}, the final training objective is formulated as follows: $\mathcal{L} = \mathcal{L}_{diff} + \mathcal{L}_{prior} + \mathcal{L}_{align}$.
\section{Experiment}

\begin{table*}
	\centering
	\small
	\resizebox{0.95\textwidth}{!}{
	\begin{tabular}{lcc|cc|ccc}
		\toprule
		                 & \multicolumn{2}{c}{\textbf{LibriTTS}} & \multicolumn{2}{c}{\textbf{VCTK}}     & \multicolumn{3}{c}{\textbf{Training Dataset}}  \\
		{\textbf{Model}} & SECS ($\uparrow$)       & CER ($\downarrow$)      & SECS ($\uparrow$)   & CER  ($\downarrow$)  & \texttt{clean-100} & \texttt{clean-360} & VCTK\\
		\midrule[0.8pt]  
		\textbf{Ground Truth} \textit{(oracle)} & {85.13 $\pm$ 2.43} & {3.88 $\pm$ 2.32} & {82.71 $\pm$ 1.50} & {1.92 $\pm$ 2.33} & -      & -     & -  \\
		\textbf{Meta-StyleSpeech} & {84.06 $\pm$ 2.23} & {3.41 $\pm$ 1.59} & {75.09 $\pm$ 2.27} & {3.82 $\pm$ 1.29} & O      & O     & X \\
		\textbf{YourTTS}$^\dagger$ & {84.47 $\pm$ 0.87} & {5.09 $\pm$ 1.60} & {82.79 $\pm$ 1.05} & {4.34 $\pm$ 1.69} & O      & O     & O   \\
		\textbf{Grad-TTS} \textit{(any-speaker)}  & {80.53 $\pm$ 1.27} & {6.87 $\pm$ 2.23} & {67.76 $\pm$ 1.23} & {8.17 $\pm$ 2.18} & O      & X     & X  \\
		\midrule[0.5pt]
		\textbf{Grad-StyleSpeech} \textit{(clean, ours)}    & {85.86 $\pm$ 0.84} & \bf {2.79 $\pm$ 1.27} & {80.46 $\pm$ 1.03} & \bf {2.49 $\pm$ 0.95} & O     & X     & X  \\
		\textbf{Grad-StyleSpeech} \textit{(full, ours)}    & \bf {87.51 $\pm$ 0.86} & {4.12 $\pm$ 1.64} & \bf {83.64 $\pm$ 0.78} & {3.65 $\pm$ 1.46} & O     & O     & X  \\
		\bottomrule
	\end{tabular}
	}
 	\vspace{-0.075in}
	\caption{\textbf{Objective evaluation for zero-shot adaptation.} Experimental results on LibriTTS and VCTK where the models are not fine-tuned. 
	We report the training dataset for a fair comparison and $\dagger$ indicates the model used VCTK in training.}
 	\vspace{-0.15in}
 	\label{tab:objective}
\end{table*}

\begin{table}
	\centering
	\small
	\resizebox{0.49\textwidth}{!}{
	\begin{tabular}{lcc|cc}
		\toprule
		                 & \multicolumn{2}{c}{\textbf{LibriTTS}} & \multicolumn{2}{c}{\textbf{VCTK}}   \\
		{\textbf{Model}} & MOS ($\uparrow$)  & SMOS ($\uparrow$) & MOS ($\uparrow$) & SMOS ($\uparrow$) \\
		\midrule[0.8pt]  
		\textbf{Ground Truth} \textit{(oracle)} & {4.57 $\pm$ 0.13} & {3.50 $\pm$ 0.24} & {4.32 $\pm$ 0.12} & {4.06 $\pm$ 0.16} \\
		\textbf{Meta-StyleSpeech} & {2.77 $\pm$ 0.27} & {2.77 $\pm$ 0.27} & {3.59 $\pm$ 0.18} & {3.75 $\pm$ 0.17}  \\
		\textbf{YourTTS}$^\dagger$ & {2.91 $\pm$ 0.25} & {2.68 $\pm$ 0.25} & {3.24 $\pm$ 0.17} & {3.17 $\pm$ 0.17}  \\
		\textbf{Grad-TTS} \textit{(any-speaker)}  & {2.47 $\pm$ 0.20} & {2.03 $\pm$ 0.20} & {2.05 $\pm$ 0.18} & {1.69 $\pm$ 0.13}  \\
		\midrule[0.5pt]
		\textbf{Grad-StyleSpeech} \textit{(clean)}    & \bf{4.18 $\pm$ 0.18} & \bf{3.83 $\pm$ 0.25} & \bf{4.13 $\pm$ 0.14} & \bf{3.95 $\pm$ 0.16}  \\
		\bottomrule
	\end{tabular}
	}
 	\vspace{-0.075in}
	\caption{\textbf{Subjective evaluation for zero-shot adaptation.} Experimental results on LibriTTS and VCTK where the models are not fine-tuned.} 
 	\vspace{-0.05in}
 	\label{tab:subjective}
\end{table}

\subsection{Experimental Setup}

\subsubsection{Dataset}

We train our model on the multi-speaker English speech dataset LibriTTS~\cite{LibriTTS}, which contains 110 hours audio of 1,142 speakers from the audiobook recordings. 
Specifically, we use the \texttt{clean-100} and \texttt{clean-360} subsets for the training and \texttt{test-clean} for the evaluation.
We also use the VCTK~\cite{VCTK} dataset which includes 110 English speakers for evaluation of the unseen speaker adaptation capability.

\subsubsection{Implementation Details}
We stack four transformer blocks~\cite{Transformer} for the text encoder and the style-adaptive encoder, respectively.
Especially, for the style-adaptive encoder and mel-style encoder, we adopt the same architecture with Style-Adaptive Layer Normalization as in Meta-StyleSpeech~\cite{StyleSpeech}.
For aligner, we use the same architecture and loss functions from the original paper~\cite{Align}.
We use the same architecture of the U-Net and linear attention for the noise estimation network $\boldsymbol{\epsilon}_\theta$ from the Grad-TTS~\cite{GradTTS}.
We use the Maximum Likelihood SDE solver~\cite{DiffVC} for faster sampling and take 100 denoising steps.
We train our model for 1M steps with batch size 8 on single TITAN RTX GPU, Adam optimizer, and learning rate as in Meta-StyleSpeech~\cite{StyleSpeech}.

\subsubsection{Baselines}
For a fair comparison, we compare our model against recent works having the official implementation and exclude end-to-end synthesis models. We use the same audio processing setup including the sampling rate of 16khz with previous works~\cite{StyleSpeech}.
As a vocoder, we use the same HiFi-GAN~\cite{HifiGAN} for all models. 

Details for each baseline we used are described as follows:
\textbf{1) Ground Truth \textit{(oracle)}}: This is a ground truth speech.
\textbf{2) Meta-StyleSpeech}~\cite{StyleSpeech}:This is the zero-shot any-speaker adaptive TTS model that utilizes the meta-learning for better unseen speaker adaptation.
\textbf{3) YourTTS}~\cite{YourTTS}: This is the zero-shot any-speaker adaptive TTS model based on the flow-based model~\cite{Glow-TTS}. In addition, this work uses the speaker consistency loss to intend the model to generate more similar speech.
\textbf{4) Grad-TTS \textit{(any-speaker)}}~\cite{GradTTS}: This is the model that modifies the Grad-TTS into the any-speaker version. In detail, we replace the fixed speaker table with the mel-style encoder~\cite{StyleSpeech} and train all modules from the scratch.
\textbf{5) Grad-StyleSpeech} (ours): This is our proposed model with the mel-style encoder and score-based diffusion models.

\subsubsection{Evaluation Setup}
For objective evaluation, we use Speaker Embedding Cosine Similarity (SECS) and Character Error Rate (CER) metrics used in YourTTS~\cite{YourTTS}. In detail, for SECS evaluation, we use the speaker encoder from the resemblyzer repository~\footnote{https://github.com/resemble-ai/Resemblyzer}.
For CER evaluation, we use the pre-trained ASR model from NeMo framework~\footnote{https://github.com/NVIDIA/NeMo}.
We follow the same evaluation setup of YourTTS for objective evaluation.
For ground truth, we randomly sample another speech of the same speaker.
For subjective evaluation, we recruit 16 evaluators and conduct human evaluations with naturalness (Mean Opinion Score; MOS) and similarity (Similarity MOS; SMOS) measures.
For subjective evaluation, we sample 10 speakers from each test set and randomly select the ground truth and reference speech for each speaker and normalize all audios with Root Mean Square normalization to mitigate a bias from the volume. 
In fine-tuning experiments, we jointly fine-tune Grad-StyleSpeech with training data from \texttt{clean-100} dataset for a regularization.

\begin{figure}
    \centering
    \includegraphics[width=0.95\linewidth]{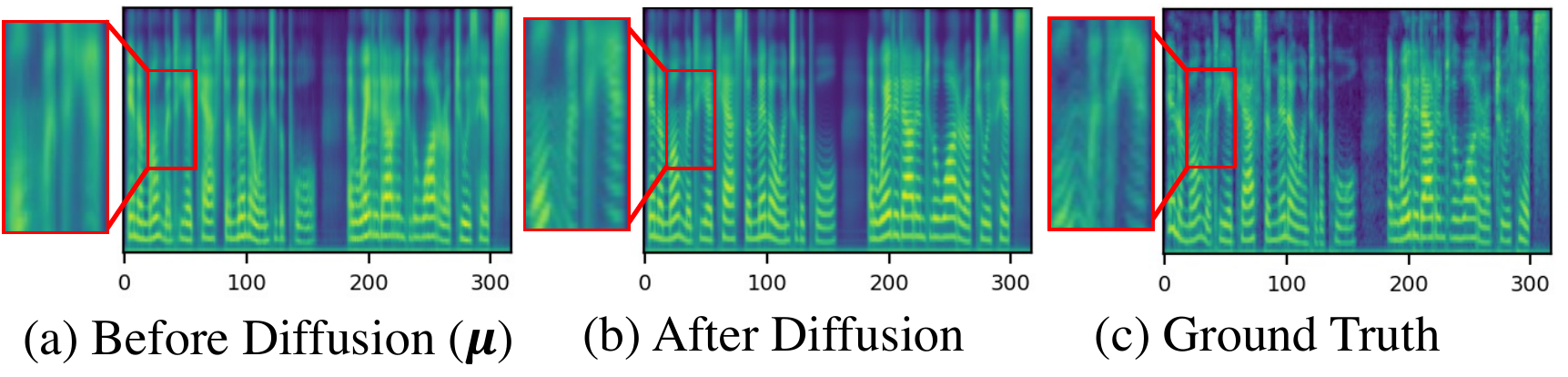}
     \vspace{-0.1in}
    \caption{\textbf{Qualitative Visualization.} We visualize the synthesized mel-spectrograms from our model (a) before the diffusion models and (b) after the diffusion models. We use the same duration with regards to the (c) ground truth speech.}
    \vspace{-0.15in}
    \label{fig:mel_comparison}
\end{figure}

\begin{figure}
    \centering
    \includegraphics[width=0.95\linewidth]{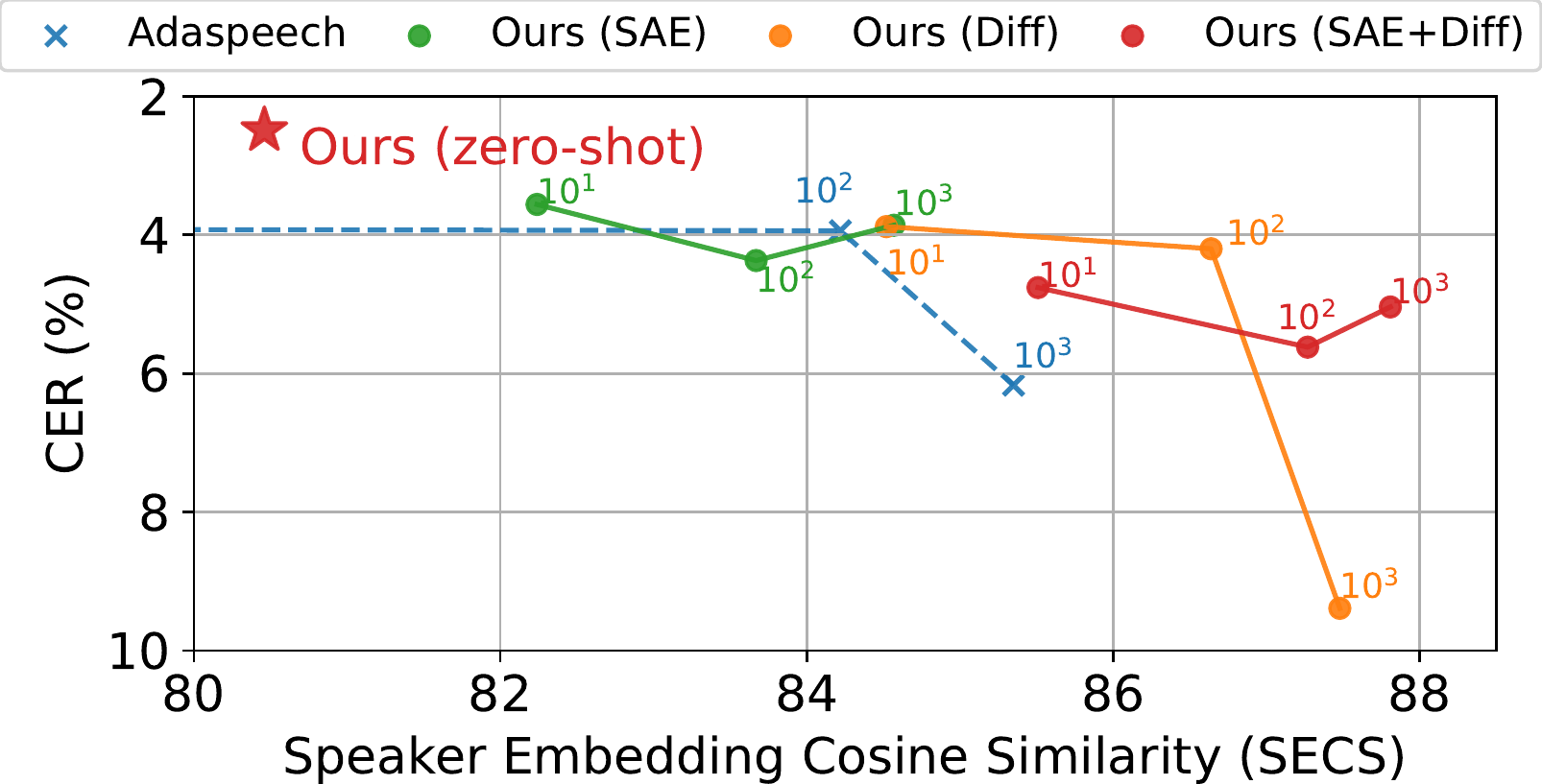}
    \vspace{-0.12in}
    \caption{\textbf{Objective evaluation for few-shot fine-tuning.}
    We plot SECS and CER varying fine-tuning steps on VCTK dataset. SAE and Diff indicate that we fine-tune the parameters of Style-Adaptive Encoder and Diffusion model, respectively.}
    \vspace{-0.15in}
    \label{fig:finetune}
\end{figure}

\vspace{-0.1in}
\subsection{Experimental Results}
In Table~\ref{tab:objective} and~\ref{tab:subjective}, we present evaluation results for zero-shot adaptation performance on unseen speakers.
In short, Grad-StyleSpeech outperforms other baselines.
Notably, our model outperforms the modified version of Grad-TTS~\cite{GradTTS}, showing that such any-speaker adaption performance is derived not only from the diffusion models but also from the hierarchical transformer encoder.
Moreover, in Table~\ref{tab:objective}, \textit{full} model outperforms others in speaker similarity measure but shows bad performance on the character error rate due to the low-quality samples in \texttt{clean-360} subset. 
We believe our method might show impressive performance if we train it on the dataset containing more clean samples from diverse speakers. 

We plot the mel-spectrograms from our model and ground truth in Figure~\ref{fig:mel_comparison} for visual analysis of generative speech. Qualitative visualization demonstrates that diffusion models help to overcome the over-smoothness issues in previous works~\cite{oversmoothness} by modeling the high-frequency components in detail.

In Figure~\ref{fig:finetune}, we plot the objective evaluation on the VCTK dataset while fine-tuning Grad-StyleSpeech on randomly selected 20 speeches of each unseen speaker. As a baseline, we use AdaSpeech~\cite{AdaSpeech} and conduct fine-tuning up to 1,000 steps.
We observe that our model outperforms AdaSpeech by fine-tuning parameters of both diffusion model and style-adaptive encoder only 100 steps of fine-tuning. 
We also conduct ablation studies by fine-tuning the selected parameters of our model. We empirically find that the fine-tuning of both the style-adaptive encoder and diffusion model is highly effective in better performance, especially for speaker similarity.

We upload audio samples used in experiments including additional samples on ESD~\cite{ESD} dataset at demo page (\href{https://nardien.github.io/grad-stylespeech-demo}{https://nardien.github.io/grad-stylespeech-demo}). Please refer to them for detailed evaluation of our method.
\section{Conclusion}
We proposed a text-to-speech synthesis model named \textit{Grad-StyleSpeech} which generates any-speaker adaptive speech with a high fidelity given the reference speech of the target speaker.
We first embed the text into the sequence of the representations conditioned on the target speaker through the hierarchical transformer encoder.
Then, we leverage the score-based diffusion models to generate a highly natural and similar speech.
Our results show that the quality of generated speech from ours highly outperforms previous baselines in both objective and subjective measures on both naturalness and similarity.

\bibliographystyle{IEEEbib}
\bibliography{strings,refs}

\begin{thebibliography}{10}

\bibitem{Transformer}
Ashish Vaswani, Noam Shazeer, Niki Parmar, Jakob Uszkoreit, Llion Jones,
  Aidan~N. Gomez, Lukasz Kaiser, and Illia Polosukhin,
\newblock ``Attention is all you need,''
\newblock in {\em NeurIPS 2017}, 2017, pp. 5998--6008.

\bibitem{Glow-TTS}
Jaehyeon Kim, Sungwon Kim, Jungil Kong, and Sungroh Yoon,
\newblock ``Glow-tts: {A} generative flow for text-to-speech via monotonic
  alignment search,''
\newblock in {\em NeurIPS}, 2020.

\bibitem{FastSpeech2}
Yi~Ren, Chenxu Hu, Xu~Tan, Tao Qin, Sheng Zhao, Zhou Zhao, and Tie{-}Yan Liu,
\newblock ``Fastspeech 2: Fast and high-quality end-to-end text to speech,''
\newblock in {\em {ICLR}}, 2021.

\bibitem{HifiGAN}
Jungil Kong, Jaehyeon Kim, and Jaekyoung Bae,
\newblock ``Hifi-gan: Generative adversarial networks for efficient and high
  fidelity speech synthesis,''
\newblock in {\em NeurIPS}, 2020.

\bibitem{DDPM}
Jonathan Ho, Ajay Jain, and Pieter Abbeel,
\newblock ``Denoising diffusion probabilistic models,''
\newblock in {\em NeurIPS}, 2020.

\bibitem{SDE}
Yang Song, Jascha Sohl{-}Dickstein, Diederik~P. Kingma, Abhishek Kumar, Stefano
  Ermon, and Ben Poole,
\newblock ``Score-based generative modeling through stochastic differential
  equations,''
\newblock in {\em {ICLR}}, 2021.

\bibitem{DiffTTS}
Myeonghun Jeong, Hyeongju Kim, Sung~Jun Cheon, Byoung~Jin Choi, and Nam~Soo
  Kim,
\newblock ``Diff-tts: {A} denoising diffusion model for text-to-speech,''
\newblock in {\em Interspeech}. 2021, pp. 3605--3609, {ISCA}.

\bibitem{GradTTS}
Vadim Popov, Ivan Vovk, Vladimir Gogoryan, Tasnima Sadekova, and Mikhail~A.
  Kudinov,
\newblock ``Grad-tts: {A} diffusion probabilistic model for text-to-speech,''
\newblock in {\em {ICML}}, 2021.

\bibitem{Guided-TTS}
Heeseung Kim, Sungwon Kim, and Sungroh Yoon,
\newblock ``Guided-tts: {A} diffusion model for text-to-speech via classifier
  guidance,''
\newblock in {\em {ICML}}, 2022, pp. 11119--11133.

\bibitem{VITS}
Jaehyeon Kim, Jungil Kong, and Juhee Son,
\newblock ``Conditional variational autoencoder with adversarial learning for
  end-to-end text-to-speech,''
\newblock in {\em {ICML}}, 2021, pp. 5530--5540.

\bibitem{MultiSpeech}
Mingjian Chen, Xu~Tan, Yi~Ren, Jin Xu, Hao Sun, Sheng Zhao, and Tao Qin,
\newblock ``Multispeech: Multi-speaker text to speech with transformer,''
\newblock in {\em Interspeech}, 2020, pp. 4024--4028.

\bibitem{NeuralVoiceCloning}
Sercan~{\"{O}}mer Arik, Jitong Chen, Kainan Peng, Wei Ping, and Yanqi Zhou,
\newblock ``Neural voice cloning with a few samples,''
\newblock in {\em NeurIPS 2018}, 2018, pp. 10040--10050.

\bibitem{SEA}
Yutian Chen, Yannis~M. Assael, Brendan Shillingford, David Budden, Scott~E.
  Reed, Heiga Zen, Quan Wang, Luis~C. Cobo, Andrew Trask, Ben Laurie,
  {\c{C}}aglar G{\"{u}}l{\c{c}}ehre, A{\"{a}}ron van~den Oord, Oriol Vinyals,
  and Nando de~Freitas,
\newblock ``Sample efficient adaptive text-to-speech,''
\newblock in {\em {ICLR} 2019}. 2019, OpenReview.net.

\bibitem{SpeakerAdaptation}
Ye~Jia, Yu~Zhang, Ron~J. Weiss, Quan Wang, Jonathan Shen, Fei Ren, Zhifeng
  Chen, Patrick Nguyen, Ruoming Pang, Ignacio Lopez{-}Moreno, and Yonghui Wu,
\newblock ``Transfer learning from speaker verification to multispeaker
  text-to-speech synthesis,''
\newblock in {\em NeurIPS}, 2018, pp. 4485--4495.

\bibitem{StyleSpeech}
Dongchan Min, Dong~Bok Lee, Eunho Yang, and Sung~Ju Hwang,
\newblock ``Meta-stylespeech : Multi-speaker adaptive text-to-speech
  generation,''
\newblock in {\em {ICML}}, 2021.

\bibitem{AdaSpeech}
Mingjian Chen, Xu~Tan, Bohan Li, Yanqing Liu, Tao Qin, Sheng Zhao, and
  Tie{-}Yan Liu,
\newblock ``Adaspeech: Adaptive text to speech for custom voice,''
\newblock in {\em {ICLR} 2021}. 2021, OpenReview.net.

\bibitem{YourTTS}
Edresson Casanova, Julian Weber, Christopher~Dane Shulby, Arnaldo~C{\^{a}}ndido
  J{\'{u}}nior, Eren G{\"{o}}lge, and Moacir~A. Ponti,
\newblock ``Yourtts: Towards zero-shot multi-speaker {TTS} and zero-shot voice
  conversion for everyone,''
\newblock in {\em {ICML}}, 2022.

\bibitem{oversmoothness}
Yi~Ren, Xu~Tan, Tao Qin, Zhou Zhao, and Tie{-}Yan Liu,
\newblock ``Revisiting over-smoothness in text to speech,''
\newblock in {\em {ACL}}, 2022, pp. 8197--8213.

\bibitem{Anderson}
Brian~D.O. Anderson,
\newblock ``Reverse-time diffusion equation models,''
\newblock {\em Stochastic Processes and their Applications}, vol. 12, no. 3,
  pp. 313--326, 1982.

\bibitem{Align}
Rohan Badlani, Adrian Lancucki, Kevin~J. Shih, Rafael Valle, Wei Ping, and
  Bryan Catanzaro,
\newblock ``One {TTS} alignment to rule them all,''
\newblock in {\em {ICASSP}}, 2022, pp. 6092--6096.

\bibitem{LibriTTS}
Heiga Zen, Viet Dang, Rob Clark, Yu~Zhang, Ron~J. Weiss, Ye~Jia, Zhifeng Chen,
  and Yonghui Wu,
\newblock ``Libritts: {A} corpus derived from librispeech for text-to-speech,''
\newblock in {\em Interspeech}, 2019, pp. 1526--1530.

\bibitem{VCTK}
Junichi Yamagishi, Christophe Veaux, and Kirsten MacDonald,
\newblock ``Cstr vctk corpus: English multi-speaker corpus for cstr voice
  cloning toolkit (version 0.92),''
\newblock 2019.

\bibitem{DiffVC}
Vadim Popov, Ivan Vovk, Vladimir Gogoryan, Tasnima Sadekova, Mikhail~Sergeevich
  Kudinov, and Jiansheng Wei,
\newblock ``Diffusion-based voice conversion with fast maximum likelihood
  sampling scheme,''
\newblock in {\em ICLR}, 2022.

\bibitem{ESD}
Kun Zhou, Berrak Sisman, Rui Liu, and Haizhou Li,
\newblock ``Emotional voice conversion: Theory, databases and {ESD},''
\newblock {\em Speech Commun.}, vol. 137, pp. 1--18, 2022.

\end{thebibliography}

\end{document}